\documentclass[
	aps, pra,
        superscriptaddress,
        twocolumn,
	10pt,
	floatfix, 
	floatfix,nobibnotes,
        nofootinbib,
	tightenlines
]{revtex4-2}
\pdfoutput=1
\usepackage[final]{graphicx}
\usepackage{times,bbm,amsmath,amssymb}
\usepackage{epsfig,color}
\usepackage{xcolor}
\usepackage{hyperref}
\hypersetup{
    colorlinks = true
}
\usepackage{cleveref}
\usepackage{microtype}

\usepackage{booktabs}
\usepackage{float}
\usepackage[caption = false]{subfig}

\usepackage[greek,english]{babel}
\usepackage{thumbpdf,enumerate}
\usepackage{booktabs}
\usepackage{sidecap}
\usepackage{pst-grad,bm}
\usepackage{epigraph}
\usepackage{longtable}
\usepackage{ulem}
\usepackage{acronym}
\usepackage{physics}
\usepackage{dsfont}
\usepackage{tabularx}
\usepackage{mathtools}
\usepackage{physics}

\usepackage{easyReview}
\usepackage{hyperref}
\usepackage{comment}
\begin{document}

\title{Boson Sampling with a reconfigurable 128 modes 3D integrated photonic circuit}

\author{Simone Di Micco}
\affiliation{Dipartimento di Fisica - Sapienza Universit\`{a} di Roma, P.le Aldo Moro 5, I-00185 Roma, Italy}

\author{Francesco Hoch}
\affiliation{Dipartimento di Fisica - Sapienza Universit\`{a} di Roma, P.le Aldo Moro 5, I-00185 Roma, Italy}

\author{Alessandro Ciorra}
\affiliation{Dipartimento di Fisica - Sapienza Universit\`{a} di Roma, P.le Aldo Moro 5, I-00185 Roma, Italy}

\author{Daniel Carvalho de Salles}
\affiliation{Dipartimento di Fisica - Sapienza Universit\`{a} di Roma, P.le Aldo Moro 5, I-00185 Roma, Italy}

\author{Nicol\`{o} Spagnolo}
\affiliation{Dipartimento di Fisica - Sapienza Universit\`{a} di Roma, P.le Aldo Moro 5, I-00185 Roma, Italy}

\author{Taira Giordani}
\affiliation{Dipartimento di Fisica - Sapienza Universit\`{a} di Roma, P.le Aldo Moro 5, I-00185 Roma, Italy}

\author{Gonzalo Carvacho}
\affiliation{Dipartimento di Fisica - Sapienza Universit\`{a} di Roma, P.le Aldo Moro 5, I-00185 Roma, Italy}

\author{Niki Di Giano}
\affiliation{Dipartimento di Fisica - Politecnico di Milano, Piazza Leonardo da Vinci, 32, 20133 Milano, Italy}
\affiliation{Istituto di Fotonica e Nanotecnologie, Consiglio Nazionale delle Ricerche (IFN-CNR), 
Piazza Leonardo da Vinci, 32, 20133 Milano, Italy}

\author{Marco Gardina}
\affiliation{Ephos, Viale Decumano 34, 20157 Milano, Italy}

\author{Andrea Crespi}
\affiliation{Dipartimento di Fisica - Politecnico di Milano, Piazza Leonardo da Vinci, 32, 20133 Milano, Italy}
\affiliation{Istituto di Fotonica e Nanotecnologie, Consiglio Nazionale delle Ricerche (IFN-CNR), 
Piazza Leonardo da Vinci, 32, 20133 Milano, Italy}

\author{Francesco Ceccarelli}
\affiliation{Ephos, Viale Decumano 34, 20157 Milano, Italy}
\affiliation{Istituto di Fotonica e Nanotecnologie, Consiglio Nazionale delle Ricerche (IFN-CNR), 
Piazza Leonardo da Vinci, 32, 20133 Milano, Italy}

\author{Roberto Osellame}
\affiliation{Ephos, Viale Decumano 34, 20157 Milano, Italy}
\affiliation{Istituto di Fotonica e Nanotecnologie, Consiglio Nazionale delle Ricerche (IFN-CNR), 
Piazza Leonardo da Vinci, 32, 20133 Milano, Italy}

\author{Fabio Sciarrino}
\affiliation{Dipartimento di Fisica - Sapienza Universit\`{a} di Roma, P.le Aldo Moro 5, I-00185 Roma, Italy}

\begin{abstract}
\textbf{ Integrated quantum photonics has emerged as one of the leading platforms for scaling quantum information processing, offering compact, stable, and low-loss hardware with precise phase and mode control.
{A}dvances in integrated photonics architectures and active programmability now enable complex, high-dimensional transformations essential for quantum advantage tasks. We introduce an integrated, reconfigurable 3D photonic device with 128 modes for manipulation of single-photon quantum states (Qolossus 3D). Leveraging a continuously coupled architecture and thermo-optic
programmability, the platform implements reconfigurable unitary transformations at unprecedented scale for integrated quantum optics. Exploiting indistinguishable single photons demultiplexed from a quantum dot source, we perform Boson Sampling across the large-dimensional chip and analyse the resulting output distributions for up to 4 photons.  We then exploit it to demonstrate
randomness generation via Boson Sampling. Agreement with theoretical predictions validates both the device’s reconfigurable operation and the generation of random numbers. Our results highlight the scalability, stability, and precise control of integrated photonics for quantum information processing.}
\end{abstract}

\maketitle

%%%%%%%%%%%%%%%%%%%%%%%%%%%%%%%%%%%%%%%%%%%%%%%%%%%%%%%%%%%%%%%%%%%%%%%%%%%%%%%%%%%%%%%%%%%%%%%%%%
\section{Introduction}

Integrated photonic quantum technologies have emerged as promising platforms for the implementation of quantum algorithms \cite{Wang2020, Pelucchi2021, Giordani2023}. Compared to free-space optics, integrated photonic circuits offer compactness, stability, and full programmability, while enabling precise control over photonic modes and their interactions. In this context, Boson Sampling \cite{10.1145/1993636.1993682, 10.1117/1.AP.1.3.034001}, a computational problem tied to the quantum interference of indistinguishable photons, has served as a foundational model for the development of photonic quantum algorithms \cite{Huh2015_vibronic, Bromley_2020, chabaud2021quantum}. The computational hardness of Boson Sampling arises from the task of drawing samples from the output distribution generated by the interference of indistinguishable photons in a linear optical evolution configured according to a Haar-random \cite{Edelman_Rao_2005}
transformation \cite{10.1145/1993636.1993682}. This complexity is fundamentally rooted in the exponential scaling of the permanent calculation required to predict the output probabilities. Early experimental demonstrations have been extensively explored in planar integrated photonic architectures, ranging from bulk-optical setups to two-dimensional integrated photonic  devices \cite{doi:10.1126/science.1231440, doi:10.1126/science.1231692, Crespi2013, Tillmann2013, PhysRevLett.118.130503, PhysRevLett.118.190501, Wang2017, PhysRevLett.121.250505}. These planar implementations have demonstrated the feasibility of on-chip quantum interference experiments and established the foundation for scalable quantum optical implementations, with progressively increasing numbers of photons and modes being successfully demonstrated across various material platforms including silicon, silicon nitride, and lithium niobate.

Recent developments in three-dimensional (3D) photonic circuits, continuously-coupled waveguide arrays, and integrated interferometers have further expanded the landscape of quantum optical devices, enabling high-dimensional and scalable implementations \cite{Arriola:13, Hoch2022, Zhou2024, Youssry2024,Yang2024,Yang2025}. These advances have facilitated the first experimental realisations of Boson Sampling in small-scale 3D integrated circuits, typically employing probabilistic parametric down-conversion-based photon sources. This intrinsic complexity has established Boson Sampling not only as a testbed for quantum advantage \cite{zhong2020quantum, zhong2021phase, madsen2022, Deng2023, liu2025robustq} but has also opened new avenues for exploring the computational capabilities of 3D photonic architectures in quantum information processing.

Beyond quantum computational advantage demonstrations, Boson Sampling has emerged as a valuable tool in diverse applications, including manipulation of random variables \cite{Patel2019, Hoch2024, Rodari2024}, acceleration of Monte Carlo simulation \cite{anguita2025_MC}, and, very recently, the generation of certified random numbers \cite{Shi, 10.1117/12.3053219}. Random number generation plays a critical role in numerous applications, ranging from cryptographic protocols \cite{RevModPhys.74.145, Pirandola:20, RevModPhys.94.025008} and statistical sampling \cite{randsampling} to machine learning and quantum simulation \cite{MLSYS2022_427e0e88, randomnesneural}. While quantum systems offer intrinsic unpredictability and integrated photonic platforms have been instrumental in realizing high-performance quantum random number generators \cite{Li:24, regazzoni2021highspeedintegratedquantum, Abellan:16, Paraïso2021, 10.1063/5.0056027, bruynsteen_100_2023,Marangon2024}, the utilization of Boson Sampling for
randomness generation remains largely underexplored, with experimental realizations \cite{Shi} confined to small-scale, low-photon-number demonstrations using probabilistic parametric sources.

\begin{figure}[t]
    \centering
    \includegraphics[width=\linewidth]{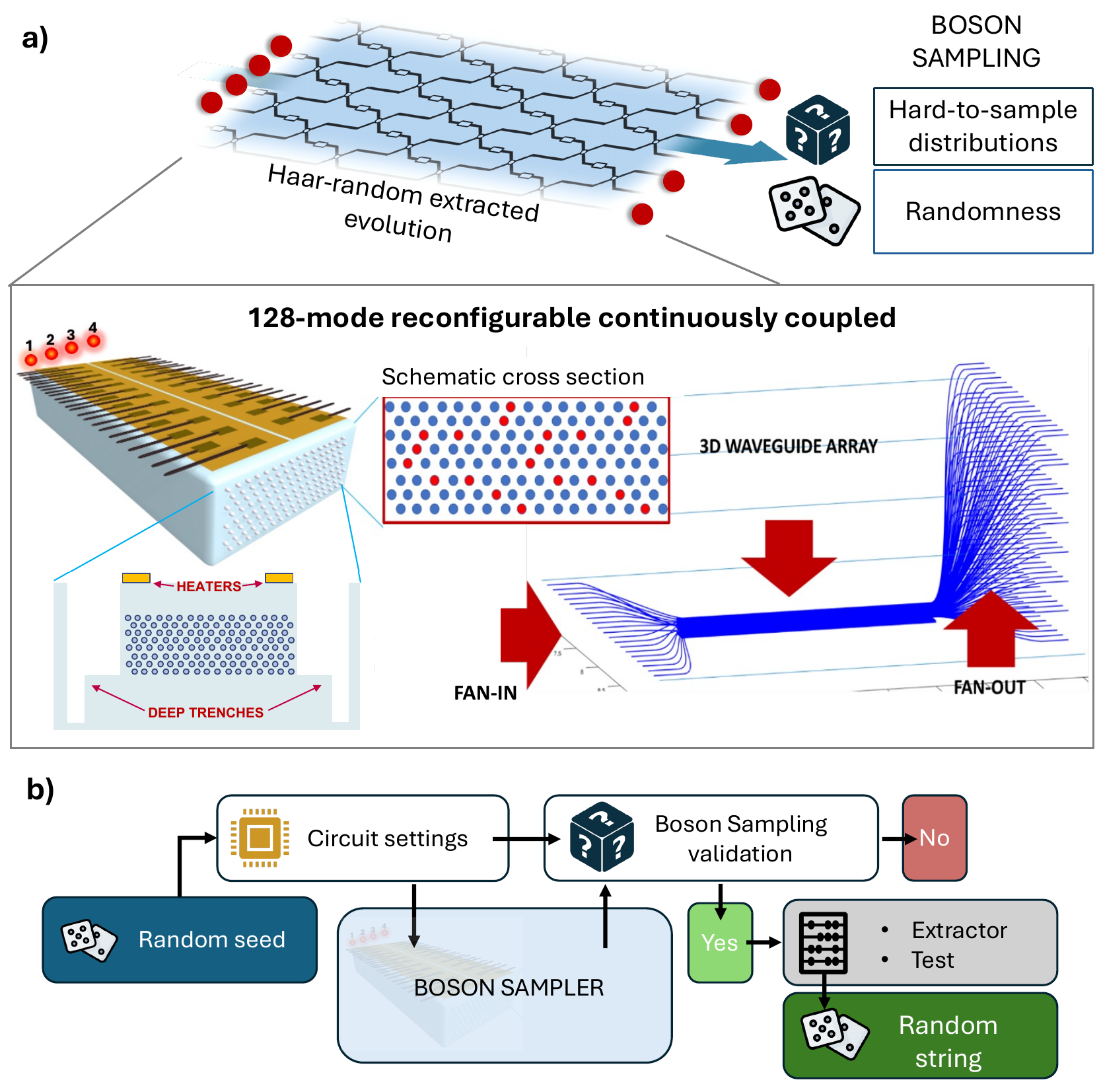}
    \caption{\textbf{High-dimensional Boson Samplers within a 3D photonic chip.} \textbf{a)} Boson Sampling exploits multi-photon interference in a randomly extracted large-scale optical transformation to perform a task that is computationally intractable for classical systems. The programmable high-dimensional integrated optical circuit is realised through the femtosecond laser writing technique and comprises 128 optical modes arranged in a triangular lattice (see inset). In red, we have highlighted the 20 modes connected to the input fiber array. The transformation $U$ can be varied by harnessing the thermo-optic effect, and 24 thermal phase shifters are realised to this purpose above the waveguide array, exploiting deep trenches in the substrate to increase the tuning efficiency. 
    \textbf{b)} The developed scheme has been adopted to carry out Boson Sampling experiments, to validate them and finally to generate strings of random bits.}
    \label{fig:scheme}
\end{figure}

In this work, we demonstrate Boson Sampling using a large reconfigurable 3D photonic architecture (Fig.~\ref{fig:scheme}). Our work demonstrates Boson Sampling on a continuously-coupled, reconfigurable three-dimensional integrated photonic processor with 128 spatial modes, combining (i) large-scale on-chip programmability via thermo-optic control, (ii) an on-demand, high-purity quantum-dot single-photon source with active time-to-spatial demultiplexing, and (iii) a custom multiplexed detection and post-processing pipeline to produce and certify random bitstrings. This integration - unprecedented in mode count for 3D femtosecond-laser-written circuits - enables repeated sampling from near-Haar-like unitaries at a scale and determinism not achievable with prior small-scale, parametric-down-conversion demonstrations \cite{Arriola:13, Hoch2022, 10.1145/1993636.1993682, Shi, 10.1117/12.3053219}. Together, these advances provide a practical, complexity-grounded route to 
establish a versatile platform for scalable photonic quantum information processing.

%%%%%%%%%%%%%%%%%%%%%%%%%%%%%%%%%%%%%%%%%%%%%%%%%%%%%%%%%%%%%%%%%%%%%%%%%%%%%%%%%%%%%%%%%%%%%%%%%%

\section{Hybrid integrated photonics platform}

\begin{figure*}[ht!]
\centering
\includegraphics[width=1\textwidth]{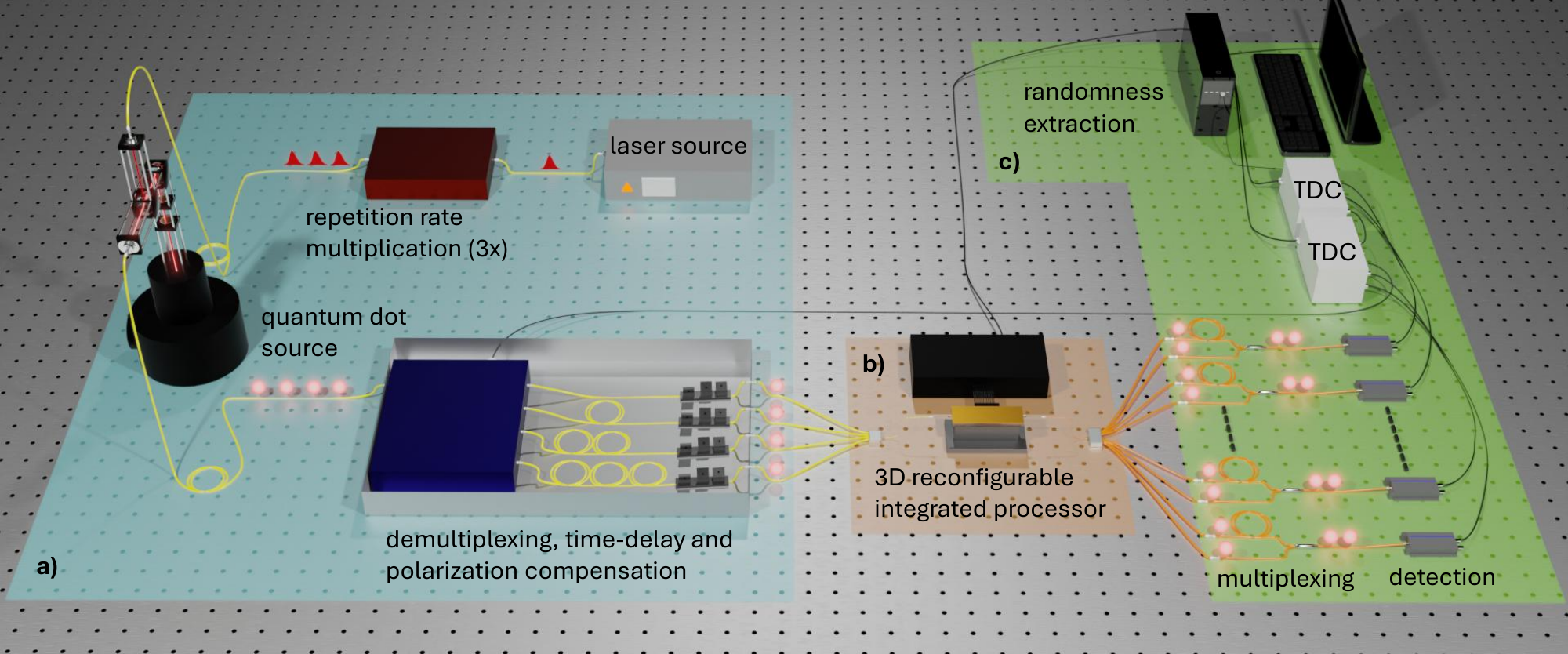}
\caption{\textbf{Experimental setup Qolossus 3D}. \textbf{a) On-demand photon generation. }We employ a Quantum Dot (QD) single-photon source housed in a cryostation maintained at 4 K, operated under resonant excitation. The emitted stream of single photons is directed to a time-to-space acousto-optic demultiplexing system, which routes the photons into four separate spatial paths. Temporal synchronisation across these paths is achieved by combining in-fiber delay loops and free-space delays for fine-tuning. \textbf{b) Evolution.} The resulting photon resource states are then injected into the 128 mode 3D integrated photonic circuit. The induced transformation can be modified via a set of micro-heaters that implement a vector of dissipated powers $\mathbf{P}$.
\textbf{c) Measurement and data analysis.} At the output of the photonic circuit, a spatial time multiplexing scheme combines pairs of optical modes, effectively reducing the number of spatial channels from 128 to 64 while doubling the temporal modes for each channel. Experimentally 54 detectors have been exploited, leading to measurements carried out over 108 modes. Photon detection is carried out using avalanche photodiodes, with timing information acquired through two synchronised 32-channel time-to-digital converters which record multifold coincidence events. The events are processed through a Von-Neumann randomness extraction protocol, which removes bias and ensures that only statistically fair bit-strings are retained. Finally, the extracted unbiased bits are fed into a SHA-256 cryptographic hash function, producing random numbers tied to the certification of Boson Sampling.}

\label{fig:schematico}

\end{figure*}

We design and implement a state-of-the-art hybrid photonic architecture, denominated \textbf{Qolossus 3D}, to enable multi-photon experiments within a large set of on-chip spatial modes. The key stages responsible for the on-demand photon generation, evolution, as well as detection and post-processing elaboration, are depicted in  Fig.~\ref{fig:schematico}.

\textit{3D programmable photonic circuits.}
The core of the developed platform is a programmable three-dimensional integrated
photonic circuit comprising $m = 128$ continuously coupled single-mode waveguides, arranged in an $8 \times 16$ triangular-lattice geometry (similarly to Ref.\cite{Hoch2022}).  In this architecture, the overall unitary operation is determined by the propagation-constant and coupling-coefficient profiles along the chip, which define a certain interaction Hamiltonian per each cross-section. The resulting transformation is the integral of these Hamiltonians over the device length. Random modulation of waveguide positions along the propagation direction induces the randomness in the final transformation, which is required in the Boson Sampling scenario. 

Fabricated in borosilicate glass by femtosecond-laser writing, the circuit is equipped with 24 resistive micro-heaters, laser patterned on a thin gold film, which can induce temperature gradients in the substrate with different spatial distributions. The temperature gradients result in refractive index gradients due to the thermo-optic effect, which in turn locally modulate the propagation constants of the waveguides and their effective coupling, thus modifying the unitary transformation implemented by the circuit. Deep trenches, processed through femtosecond laser ablation, are positioned at the sides of the circuit to improve phase shifter tuning efficiency (Fig. \ref{fig:scheme}a) by preventing heat dissipation in areas where temperature gradients are unnecessary. This continuously-coupled interferometer architecture is different from more conventional linear-optical implementations, where an arbitrary unitary transformation is produced by controlled phase shifters and balanced beam splitters in cascade \cite{10.1117/1.AP.1.3.034001}. Our 3D layout reduces routing complexity, and improves compactness and loss performance compared with discretely-coupled interferometers, facilitating stable operation over 128 modes with manageable insertion loss. Our device further shows the compatibility of thermo-optic phase shifting with large-size three-dimensional, continuously-coupled waveguide arrays. 
In the experiment presented here, we exploit up to 17 phase shifters, actuated simultaneously, to actively control the transformation $U$ produced by the device.
The interferometric evolution implemented by the chip under a heater-power vector $\mathbf{P}$ is described by a $128\times128$ unitary matrix:
$
\label{eq:U}
U(\mathbf{P})=\mathcal{T}\exp\!\left[-\,i\!\int_0^{L} H\!\left(z;\mathbf{P}\right)\,\mathrm{d}z\right],
$
where $H(z;\mathbf{P})$ is an effective coupling Hamiltonian set by the 3D lattice and the thermo-optic shifts. In the operating range, the heaters act predominantly as phase shifters producing different contributions on different sites.

\begin{figure*}[t!]
\begin{center}
\includegraphics[width=1\textwidth]{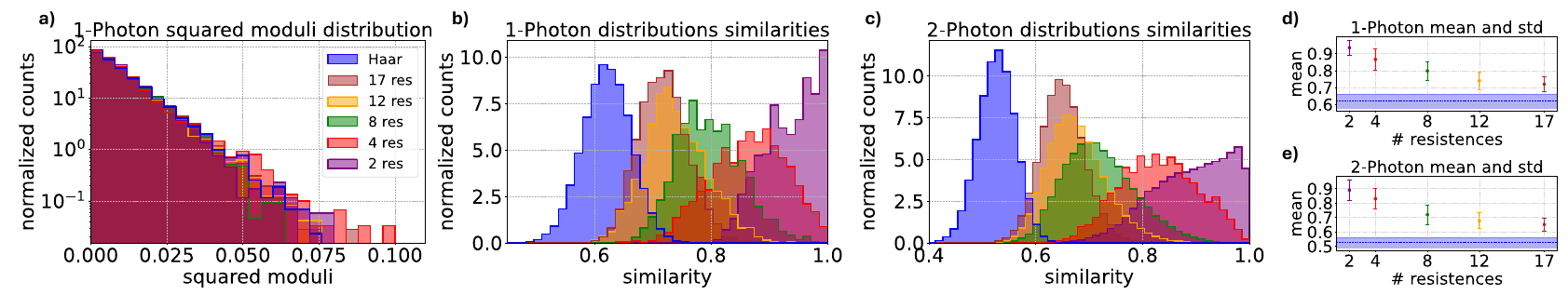}
\end{center}
\caption{
\textbf{Reconfigurability of the 3D integrated device and Haar-random sampling}. \textbf{a)} Squared moduli distribution of the matrix {elements} $U_{ij}(\textbf{P})$ when varying the currents applied to different numbers of heaters (from 2 up to 17). The experimental data are obtained by injecting single photons in a given input mode and recording the output photon distributions. The {expected} histogram corresponds to the numerical simulation with 100 matrices extracted according to the Haar-random measure. \textbf{b)} Distributions of the similarities between pairs of output vectors, where each vector represents the output distributions obtained for a given current setting \textbf{P} with 1-photon as input. We compare the distributions with the one obtained for simulated Haar-random extracted matrices. \textbf{c)} Distribution{s} of the similarities for the two-photon  {probabilities} by using different numbers of heaters. \textbf{d-e)}  {Averages} and standard deviations of the distribution for the 1-photon and 2-photon scenarios, respectively, for the different number of active {heaters}. The horizontal blue band represents the mean of the Haar random distribution, within a standard deviation. All the data are restricted to the subset of 108 detected output modes.}
\label{fig:haar}
\end{figure*}

\textit{Experimental setup.} 
A Quantum Dot single-photon source generates a stream of indistinguishable single-photons, with measured single-photon purity $g^{(2)}(0)\sim 4\%$, and a Hong-Ou-Mandel visibility $V_\mathrm{HOM}\sim83\%$. The stream of single photons is sent to a time-to-space demultiplexing system working through acousto-optic modulation, and routing sequentially streams of photons to different fiber-optic paths. Delay lines in each path ensure 
time synchronisation before the photons enter the integrated device. In addition, 
indistinguishability of the photons in the different paths is warranted by in-line polarization controllers at each path. The fiber outputs of this demultiplexing and compensation system are coupled to a selected subset of the 20 available input ports of the photonic chip. These correspond to 20 fan-in waveguides connected to specific modes of the 128 available in the 3D array (these modes are marked in red in the schematic cross-section of Fig. \ref{fig:scheme}a).  

Inside the 3D integrated photonic device, the optical modes
undergo the unitary evolution defined by the set of applied currents. At the output, a fan-out waveguide array (Fig.~\ref{fig:scheme}-\textbf{c}) connects the 128 single-mode channels to a custom multiplexing system able to combine pairs of spatial modes into distinct time bins of a single spatial mode,
by applying a delay to one of the two (see Fig. \ref{fig:schematico}). In this way, the number of required detectors is halved. In our experiments, we employ 54 detectors to perform measurements on 108 modes. In particular, single-photon  
detection is carried out using single-photon avalanche photo-diodes (SPADs), whose electrical outputs are then acquired via two 32-channel time-to-digital converters (TDC), finally connected to a computer for further data processing. 

\section{3D chip characterization}

\begin{figure*}[t]
    \centering
    \includegraphics[width=\textwidth]{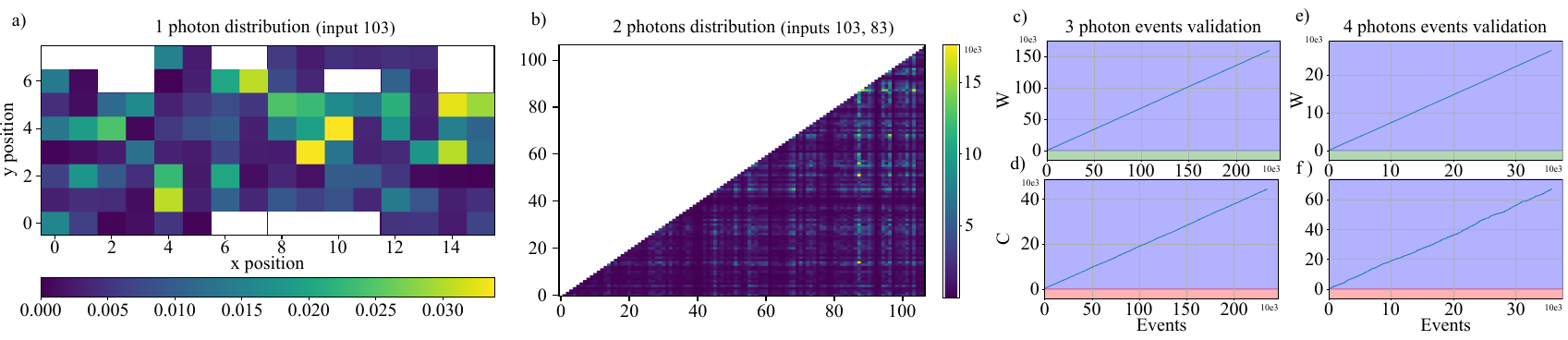}
    \caption{
    \textbf{Validation of Boson Sampling data.} Sequential likelihood–ratio tests on three-fold and four-fold, collision-free events collected at the output of the 128-mode device. {In \textbf{(a)} and \textbf{(b)}, examples of the one-photon and two-photon distributions used for the reconstruction of the unitary matrix. Panel (a) highlights in white the 20 output ports that were not measured in the experiment.}
    \textbf{(c)} \textbf{(e)} Counter $W_k$ for the uniform-sampler null hypothesis $H_{\mathrm{uni}}$: the linear growth with the number of events $k$ leaves the shaded acceptance band, thus rejecting $H_{\mathrm{uni}}$. \textbf{(d)} and \textbf{(f)} Counter $C_k$ for the fully distinguishable-photon model $H_{\mathrm{dist}}$: the monotonic increase beyond the red acceptance band excludes $H_{\mathrm{dist}}$. Horizontal axes show the cumulative number of events; vertical axes report the counter values $W_k$ and $C_k$. Procedures follow the sequential likelihood-ratio tests in \cite{Aaronson2014} and \cite{Spagnolo2014}.
    }
    \label{fig:validation}
\end{figure*}

The main goal is to investigate {the degree of reconfigurability of the device achievable by the available} thermo-optic phase shifters
driven by {the corresponding dissipated powers} \textbf{P}. We describe the induced transformation on 128 modes with the matrix  $
U_{ij}(\textbf{P})=\rho_{ij}(\textbf{P})\,e^{i\theta_{ij}(\textbf{P})}$, with $\rho_{ij}(\textbf{P})\ge0$ and  $\theta_{ij}(\textbf{P})\in(-\pi,\pi]$. As a reference one should take into account that a universal optical {device with a discretized layout} acting on 128 modes would require roughly $10^5$ parameters for a full control.

We evaluate how closely the accessible transformations approximate Haar‑random behavior - a desirable property for several quantum information applications, ranging from Boson Sampling complexity \cite{10.1145/1993636.1993682}, quantum random walks, quantum simulation to benchmarking of noisy intermediate-scale quantum photonic processors \cite{Elben2022,Lin25} - {as a function of} the number of employed thermo-optic phase shifters
(Fig.~\ref{fig:haar}). {In particular, we tested the behavior of the $U_{ij}(\textbf{P})$ when} the \textbf{P} values are obtained by uniformly sampling the electrical powers dissipated in the resistive heaters, given that the induced phase change is proportional to the power.

We carried out three different tests.
First, for a growing number of active heaters (from 2 to 17), we compare {in Fig.~\ref{fig:haar}-\textbf{a}} the measured distributions of {of the square moduli} ${\rho_{ij}^2(\textbf{P}})$ with the theoretical distributions for matrices drawn from Haar‑random unitaries. This test shows a very good agreement with Haar‑random predictions, largely independent of the degree of reconfigurability (i.e., the number of active micro-resistors
). 

Subsequent tests were designed to provide deeper insight into the device reconfigurability,
and to demonstrate its capability to sample unitary transformations that closely approximate the Haar-random distribution.
In the second study, for different configurations of the dissipated powers \textbf{P}, we record single‑photon output statistics to reconstruct an estimate of the output photon number distribution, corresponding to the square moduli of a matrix column. Since each column can be seen as a discrete probability function distribution $D_i$, then we computed the pairwise similarity between each couple of columns $D_i$ and $D'_i$  defined as $ S =\left(\sum_i  \sqrt{D_{i}D'_{i}}\right)^2 $. We then compare the obtained distribution of pairwise similarities with the corresponding theoretical distribution for columns of Haar‑random unitaries (see Fig.~\ref{fig:haar}-\textbf{b}). 
This second test highlights a dependence on the number of active micro-heaters, indicating that increased reconfigurability drives the device closer to an ideal Haar‑like behavior.

As a third test, we inject photons into pairs of input modes, and compare in Fig.~\ref{fig:haar}-\textbf{c} the similarity between the resulting output two‑fold coincidence distributions and the Haar‑random benchmark. Two-photon distributions give us further insight into the properties of the extracted transformations since they also depend on the phases $\theta_{ij}(\mathbf{P})$ of the matrices $U_{ij}(\mathbf{P})$. Again, we observe that by increasing the number of micro-heaters,
we get closer to predictions to the Haar-random benchmark.
Hence, while the 128-mode processor is not completely 
universal as full control would require $\sim 10^5$ independent parameters,
17 thermo-optic actuators already
allow us to explore a high-dimensional random
subset of the unitary group that is sufficient to reproduce Haar-like statistics on the 108 detected modes, as summarised in Fig.~\ref{fig:haar}-\textbf{d-e}.

\textit{Unitary matrix reconstruction.} As final step for the {device} calibration, we have fully characterized the transformations induced by the 128-mode photonic processor within the adopted optical input modes for a limited set of configurations  \textbf{P} \cite{superstabletomographylinearoptical}. First, we measure the output distributions when injecting single photons in a single input at a time {(Fig.~\ref{fig:validation}a)}. The outcomes give us information about the moduli $\rho_{ij}(\textbf{P})$. We then measure the {complete} two-photon output distribution when injecting two photons into pairs of distinct input modes {(Fig.~\ref{fig:validation}b)}. In this way, we obtain {also information on} the phases $\theta_{ij}(\textbf{P})$. The reconstructed submatrices (restricted to a subset of the input modes) will be exploited to validate the multi-photon experiments described in the next section. {In particular we reconstruct the sub matrices accessible with 4 input modes.}

\section{Boson Sampling validation}

\textbf{Boson Sampling validation.}
After characterizing the device, we have performed 3- and 4-photon Boson Sampling experiments with the platform described above. We now illustrate the analysis of the experimental data collected in 3‑ and 4‑photon Boson Sampling routines. In this context, data validation is crucial for assessing the correctness of the sampling process, especially in regimes where the experimental output cannot be efficiently reproduced with classical resources. In recent years, several tests have been developed to rule out classical models - such as uniform and distinguishable particle samplers - that can reproduce some features of the Boson Sampling output distribution \cite{10.1117/1.AP.1.3.034001}.
We validate the origin of the raw data as genuine multiphoton interference on the 128-mode interferometer. Specifically, thanks to the reconstructed unitary submatrices described before, we evaluate fast counters that discriminate Boson Sampling statistics from two null hypotheses: a uniform sampler \cite{Aaronson2014} and a fully distinguishable-photon model \cite{Spagnolo2014}.
Multi-photon coincidences were recorded and used to validate the Boson Sampling. The presence of non-trivial bosonic correlations in the generated events is confirmed by consistently positive log-likelihood ratios (Fig.~\ref{fig:validation}). These ratios, tested against the alternative hypotheses (uniform sampler $H_{uni}$ and distinguishable photons sampler $H_{dist}$), were obtained using the $W_k$ \cite{Aaronson2014} and $C_k$ \cite{Spagnolo2014} estimators evaluated at the $k$-th registered event. 

\textbf{Randomness extraction, hashing, and validation.}
As final step, we have then used as a proof-of-principle our platform to extract
random strings obtained via the Boson Sampling experiments. The Boson Sampling runs on the 128-mode device yield, for each trial, a binary occupancy vector $\mathbf{b}^{(t)}=(b^{(t)}_{1},\ldots,b^{(t)}_{128})$ that records whether each output mode clicked within the acquisition window. Aggregating $T$ trials forms a $T\times128$ matrix of raw bits which constitutes our entropy source. Since click probabilities can be mode-dependent (loss, detector efficiency, and coupling asymmetries), the raw matrix is not guaranteed to be unbiased or independent across rows/columns; the following post-processing converts it into uniformly distributed bits suitable for downstream applications. In particular, to remove potential first-order bias, we apply the Von Neumann (VN) extractor to each mode independently, which works in the following way: for each mode $j$, we consider its time-ordered sequence $S_j=\{b^{(1)}_{j},b^{(2)}_{j},\ldots,b^{(T)}_{j}\}$, hence, we consider pairwise subsequent bits $(b_j^{(2n)}, b_j^{(2n+1)})\ \text{with}\ n\in[0,\frac{T}{2}-1] $, and assign a bit 0 to the sequence $(0,1)$, a bit 1 to the sequence $(1,0)$, and we discard the sequences $(0,0)$ and $(1,1)$. After VN extraction, the unbiased bitstreams are shorter, reflecting the discarded pairs, and provide a more uniform entropy source. To validate the outcomes of the VN extractor, we evaluated the min-entropy $H_{min}$ of the concatenated bitstreams as a function of the block size.

\begin{figure}[t!]
\begin{center}
\includegraphics[width=\columnwidth]{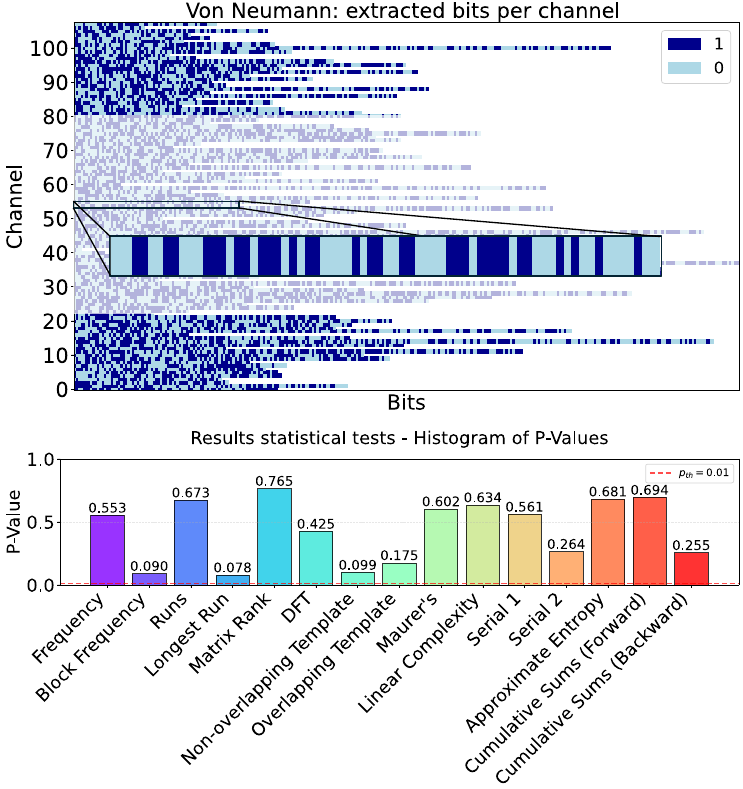}
\end{center}
\caption{\textbf{Randomness extraction and validation for 3 photons BS.} \textbf{a) Von Neumann–unbiased bitstreams}. Each row corresponds to one detector channel, showing the final unbiased bitstring obtained after the Von Neumann extractor. The length varies depending on the number of discarded pairs, reflecting statistical fluctuations in the raw data. {\textbf{b) NIST statistical test suite}. In the figure, we show the results of the NIST statistical tests for random and pseudorandom number generation for cryptographic applications. The tests certify the randomness of the generated bitstream and are considered passed for values of $p>p_{th} = 0.01$} {(dashed red line)}. {Indeed, all tests can be considered successfully passed.}}
\label{fig:bigwin}
\end{figure}

\textit{Cryptographic conditioning via hashing.}
VN unbiasing does not eliminate higher-order correlations between modes or across time. We therefore apply a strong extractor by hashing fixed-length blocks of $\tilde S$ with a preimage-resistant function (SHA-256 in our implementation). To guarantee at least 256 bits of unpredictability per digest under a device-dependent entropy model, the block length is chosen as $L \;=\; \left\lceil \frac{256}{H_{\min}} \right\rceil.$ 
Each block is mapped to a 256-bit digest; concatenating digests yields the final output stream. This conditioning acts as both entropy amplification (under the stated min-entropy bound) and a robust mixer against residual structure.

\textit{Statistical validation of the bitstream.}
The hashed stream has been finally validated via the NIST statistical test suite \cite{bassham2008statistical} which includes 15 tests: the frequency (monobit) test, frequency test within a block, the runs test, test for the longest-run-of-ones in a block, the binary matrix rank test, the Discrete Fourier Transform (DFT) test, the non-overlapping template matching test, the overlapping template matching test, Maurer's "universal statistical" test, the linear complexity test, the serial test 1, the serial test 2, the approximate entropy test, the cumulative sums test forward and the cumulative sums test backward. The tests are considered passed for a p-value above a threshold significance level $p_{th} = 0.01$. In Fig.~\ref{fig:bigwin}a we show, for a boson sampling performed with 3 photons, a subset of bitstreams after the VN extractor for each measured channel of the chip. In Fig.~\ref{fig:bigwin}b we report the results of the statistical tests. All tests give positive results, showing that the method is reliable to generate a random sequence of bits. The resulting sequences of length $\sim 3 \cdot 10^6$ bits pass standard batteries at significance $p_{th} = 0.01$ with an overall bit rate generation of $\sim 175~b/s$, establishing statistical soundness within finite-size limits. 

\section{Discussion}

Our work advances reconfigurable Fock-state platforms  by combining bright single-photon generation and active demultiplexing with a 128-mode interferometer, allowing repeated draws from ensembles of near-Haar transformations and direct access to Fock-space statistics. The Boson Sampling validation test against uniform and distinguishable-particle surrogates remain positive over $10^6$ instances for threefold coincidences, and over $10^3$ instances for fourfold coincidences, validating the presence of quantum correlations (Fig.~\ref{fig:validation}). In parallel, we map each sampling event to an occupancy bit string and implement an end-to-end randomness pipeline: per-mode von Neumann unbiasing, conservative post-VN min-entropy estimation, and cryptographic conditioning via hashing with block length set to guarantee 256 bits of unpredictability per digest. Such results validate the operation of the implemented high-dimensional, Haar-like photonic processor.

The present platform performances are bounded by source brightness and indistinguishability (setting multi-fold rates and HOM visibilities), insertion loss and detection efficiency. These factors explain residual deviations in a small subset of tests and modest drift across long time acquisitions. 
With the adoption of a brighter single-photon source and superconducting nanowires single-photon detectors currently under progress, large complexity levels of the boson sampling will be achieved increasing the correlation depth and strengthening hypothesis tests. Finally, the adoption of a larger number of active phase shifters, acting independently on the transformation performed by the waveguide array,
will enlarge the effectively sampled region of $\mathrm{U}(128)$ and then bring the
statistics closer to Haar predictions.

These results establish the present platform as a versatile foundation for quantum information applications ranging from secure communication to finance, supporting, for example, Boson Sampling-based blockchain protocols \cite{Singh2025} where unbiased entropy is critical for decentralized consensus and verifiable random beacons. Beyond direct generation, this architecture can, in principle, be exploited for randomness manipulation routines such as Quantum Bernoulli Factories \cite{dale2015provable,jiang2018quantum,Hoch2024, Rodari2024, hoch2025complexitymultifunctionalvariantsquantumtoquantum}, which are useful primitive for secure distributed tasks like blind quantum computation \cite{kashefi2017multiparty}. Additionally, the high-dimensional random evolution demonstrated here provides a resource for neuromorphic models, specifically reservoir computing \cite{Yan2024}, a neural network paradigm that exploits large-scale random evolution.

%%%%%%%%%%%%%%%%%%%%%%%%%%%%%%%%%%%%%%%%%%%%%%%%%%%%%%%%%%%%%%%%%%%%%%%%%%%%%%%%%%%%%%%%%%%%%%%%%%
\section{Acknowledgements}
The authors acknowledge support from FARE Ricerca in Italia QU-DICE Grant n. R20TRHTSPA. This work is supported by the European Union via project QLASS (“Quantum Glass-based Photonic Integrated Circuits” - Grant Agreement No. 101135876), PNRR MUR project PE0000023-NQSTI (Spoke 4) and by the ERC Advanced Grant QU-BOSS (QUantum advantage via nonlinear BOSon Sampling, grant agreement no. 884676). G.C. acknowledges support from Sapienza Grant n.
RG1241910DDF1480. 
F.C. acknowledges financial support from the project HI-LIGHT (Hybrid Integration of Laser-written Interferometers and sinGle pHoton deTectors), grant n. 2022JRSST2, funded by the Italian Ministry of University and Research (MUR) through the PRIN 2022 program (D.D. n. 104, 02/02/2022).
The photonic chip was partially fabricated at PoliFAB, the micro- and nanofabrication facility of Politecnico di Milano (\href{https://www.polifab.polimi.it/}{https://www.polifab.polimi.it/}). We acknowledge helpful discussion with Ulysse Chabaud.

%\bibliographystyle{apsrev4-1}
%\bibliography{biblio.bib}


\end{document}